# Observation of non-Hermitian many-body skin effects in Hilbert space

Weixuan Zhang[1,*], Fengxiao Di[1,*], Hao Yuan[1,*], Haiteng Wang[1], Xingen Zheng[1], Lu He[1], Houjun Sun[2], and Xiangdong Zhang[1,$]

[1] Key Laboratory of advanced optoelectronic quantum architecture and measurements of Ministry of Education, Beijing Key Laboratory of Nanophotonics & Ultrafine Optoelectronic Systems, School of Physics, Beijing Institute of Technology, 100081, Beijing, China

[2] Beijing Key Laboratory of Millimeter wave and Terahertz Techniques, School of Information and Electronics, Beijing Institute of Technology, Beijing 100081, China

*These authors contributed equally to this work. $Author to whom any correspondence should be addressed. E-mail: zhangxd@bit.edu.cn

**Non-Hermiticity greatly expands existing physical laws beyond the Hermitian framework, revealing various novel phenomena with unique properties. Up to now, most exotic non-Hermitian effects, such as exceptional points[1-8] and non-Hermitian skin effects[9-12], are discovered in single-particle systems. The interplay between non-Hermitian and many-body correlation is expected to be a more fascinating but much less explored area. Due to the complexity of the problem, current researches in this field mainly stay at the theoretical level[13-21]. The experimental observation of predicted non-Hermitian many-body phases is still a great challenging[22]. Here, we report the first experimental simulation of strongly correlated non-Hermitian many-body system, and reveal a new type of non-Hermitian many-body skin states toward effective boundaries in Hilbert space. Such an interaction-induced non-Hermitian many-body skin effect represents the aggregation of bosonic clusters with non-identical occupations in the periodic lattice. In particular, by mapping eigen-states of three correlated bosons to modes of the designed three-dimensional electric circuit, non-Hermitian many-body skin effects in Hilbert space is verified by measuring the spatial impedance response. Our finding not only discloses a new physical effect in the non-Hermitian many-body system, but also suggests a flexible platform to further investigate other non-Hermitian correlated phases in experiments.**

Exploring novel physical phenomena in strongly correlated many-body systems is one of the most difficult challenges in modern physics, where particle interactions could play key roles in the formation of various correlated quantum phases, including the ferromagnetism[23], superconductivity[24], Mott insulators[25], fractional quantum Hall states[26] and so on[27]. Up to now, especially on the experimental aspect, early efforts on the quantum many-body physics mainly focus on the Hermitian system under the assumptions of being conservative and obeying time-reversal symmetry.

Recent developments in the non-Hermitian physics have opened exciting opportunities to exhibit exotic behaviors beyond the Hermitian framework[1-12]. One representative example is the non-Hermitian skin effect[9-12], where the energy eigenvalues and corresponding eigenstates are significantly changed in a nonlocal way with distant boundary conditions. Following the discovery of non-Bloch eigenstates related to non-Hermitian skin effects, various intriguing phenomena, such as the modified bulk-boundary correspondence[9-12, 28-31] and non-Hermitian critical behavior[32], are proposed. Subsequently, the experimental observation of non-Hermitian skin effects has been realized in a variety of non-conservative systems[33-36], providing versatile platforms for exploring unconventional wave phenomena and giving rise to many novel applications in the field of the non-Hermitian sensor[37] and the topological wave funneling[38]. However, up to now, all investigations on non-Hermitian skin effects focus on single-particle systems without the translation invariance.

Comparing with the single-particle system, the non-Hermitian many-body system may show much more exotic features. Motivated by the development of non-Hermitian single-particle physics, there have been emerging theoretical interests in developing novel phenomena originated from the interplay between non-Hermitian and many-body correlations[13-21]. However, in spite of the central importance, it still remains challenging to fulfill these exotic non-Hermitian many-body phases in experiments[22]. In this case, a newly accessible and fully controllable platform is urgently expected to simulate non-Hermitian many-body systems with strong interactions. Moreover, owing to rich properties of correlated particles, finding more exotic effects without Hermitian and single-particle counterparts is also much-needed to promote the development of non-Hermitian many-body physics.

In this work, we demonstrate both in theory and experiment that the strong interaction can induce a new type of non-Hermitian skin effects in Hilbert space. Different from physical boundaries showing single-particle skin effects, the strong interaction-induced skin boundaries locate in Hilbert space of many-body systems with periodic boundary conditions. Based on the analytical derivations and numerical simulations, the appearance of interaction-induced skin effects in Hilbert space is confirmed. In experiments, using three-boson system as an illustration, we map the corresponding eigenstates to modes of the designed circuit lattice, and the non-Hermitian many-body skin effects in Hilbert space is verified by measuring the spatial impedance response. Our proposal provides a useful laboratory tool to investigate various non-Hermitian many-body systems, and may possess a great potential for the electronic signal control.

**_The theory of interaction-induced non-Hermitian many-body skin effects in Hilbert space_**. We consider correlated bosons with asymmetric hoppings on a one-dimensional (1D) chain under the periodic boundary condition. The lattice length is $L$, and the number of bosons is $N$.

In this case, the system can be described by the non-reciprocal 1D Bose-Hubbard Hamiltonian as:

$$H = -\sum_l [J^+ a^+_{l+1} a_l + J^- a^+_l a_{l+1}] + 0.5U \sum_l n_l(n_l - 1), \quad (1)$$

where $a^+_l$ ($a_l$) and $n_l = a^+_l a_l$ are the creation (annihilation) and particle number operators at the $l$th lattice site, respectively. $J^\pm$ define asymmetric hopping strengths and $U$ corresponds to the on-site interaction energy. We start to consider a simple case with $N=3$. The three-boson solution can be expanded in Hilbert space as:

$$|\psi> = \frac{1}{\sqrt{6}} \sum_{m,n,q=1}^{L} c_{mnq} a^+_m a^+_n a^+_q |0>, \quad (2)$$

where $|0>$ is the vacuum state and $c_{mnq}$ is the probability amplitude of the first boson at the site $m$, the second boson at the site $n$, and the third boson at the site $q$. Substituting Eqs. (1) and (2) into the stationary Schrödinger equation $H|\psi> = \varepsilon|\psi>$, we obtain the eigen-equation with respect to $c_{mnq}$ as:

$$\varepsilon c_{mnq} = -J^\pm (c_{m\pm1,n,q} + c_{m,n\pm1,q} + c_{m,n,q\pm1}) + U(\delta_{mn} + \delta_{mq} + \delta_{nq}) c_{mnq}. \quad (3)$$

To illustrate the distribution of probability amplitude for three bosons at different energy scales, we divide the configuration space of $c_{mnq}$ into three subspaces, as shown in Fig. 1a. Without loss of generality, the lattice length is set as $L=10$. Red dots on the diagonal line of the configuration space ($m=n=q$) correspond to the subspace with three bosons locating at the same lattice site, where the eigen-energy and dimension are $\varepsilon \sim 3U$ and 1D, as illustrated by the inset enclosed with red frame. Blue dots on the diagonal plane ($m=n\neq q$, $m=q\neq n$ and $n=q\neq m$) represent the subspace with two bosons locating at the same lattice site, as shown in the inset with blue frame, where the effective energy is $\varepsilon \sim U$ and the dimension is 2D. Black dots correspond to the subspace with three bosons locating at different lattice sites, as presented in the inset with black frame, and the corresponding energy scale and configuration dimension are $\varepsilon \sim 0$ and 3D, respectively. When the on-site interaction is extremely strong ($U \gg J^{\pm}$), the energy mismatch between these three sub-spaces becomes very large, making the higher-energy subspace perform like a potential barrier embedding in the lower-energy subspace with higher-dimensionality[42, 43]. In this case, the 2D (1D) sub-space with energy being $\varepsilon \sim U$ ($\varepsilon \sim 3U$) can be regarded as the effective boundary for the 3D (2D) sub-space with $\varepsilon \sim 0$ ($\varepsilon \sim U$).

To clarify the distribution of probability amplitude toward these effective boundaries, hopping strengths perpendicular to these boundaries should be analyzed. As shown in Fig. 1b, we plot wave vectors parallel/perpendicular to the effective 2D (and 1D) boundary in the 3D (and 2D) subspace. Here, $\boldsymbol{k}_m$, $\boldsymbol{k}_n$ and $\boldsymbol{k}_q$ represent wave vectors of three bosons in the periodic lattice. In this case, wave vectors parallel and perpendicular to the effective 2D boundary ($m = n \neq q$) in the 3D subspace can be expressed as $\boldsymbol{k}_{\parallel(m=n),\perp(m=n)} = \boldsymbol{k}_m \pm \boldsymbol{k}_n$. Moreover, $\boldsymbol{k}_{\parallel(m=n=q)}$ and $\boldsymbol{k}_{\perp(m=n=q)}$, which are in the form of $\boldsymbol{k}_{\parallel(m=n=q),\perp(m=n=q)} = \boldsymbol{k}_{\parallel(m=n)} \pm \boldsymbol{k}_q$, correspond to wave vectors parallel and perpendicular to the effective 1D boundary ($m = n = q$) in the 2D subspace, respectively.

Using these predefined wave vectors, we firstly formulate hopping amplitudes in the 3D subspace perpendicular to 2D effective boundaries. For this purpose, the three-boson Bloch Hamiltonian is expressed in the form of a Hatano-Nelson chain[39] in the direction perpendicular to the 2D boundary ($m = n \neq q$) as:

$$H(k_{\perp(m=n)}) = J_m^{\pm} e^{\pm i k_m} + J_n^{\pm} e^{\pm i k_n} = J_{\perp(m=n)}^{\pm} e^{\pm i k_{\perp(m=n)}}, \qquad (4)$$

where $J_i^{\pm}$ are the asymmetric hopping along $\pm \boldsymbol{k}_i$ and $J_{\perp(m=n)}^{\pm} = J_m^{\pm} e^{\pm i k_{\parallel(m=n)}} + J_n^{\mp} e^{\mp i k_{\parallel(m=n)}}$ correspond to the effective hopping in the direction of $\pm \boldsymbol{k}_{\perp(m=n)}$. As long as the effective hopping strengths are unbalanced, the boundary-induced localization of probability amplitudes could appear that is similar to the non-reciprocal single-particle systems with open boundaries[9-12]. Following direct analytical derivations (see S1 in the Supplementary Materials for details), the requirement for symmetric hopping amplitudes $|J^+_{\perp(m=n)}|=|J^-_{\perp(m=n)}|$ could be expressed as:

$$(J^+_m)^2 + (J^-_n)^2 = (J^-_m)^2 + (J^+_n)^2, \quad J^+_m J^-_n = J^-_m J^+_n. \tag{5}$$

Ensured by the identical principle of bosons, we note that Eq. (5) always holds, that is the balanced hopping strengths along $\pm \boldsymbol{k}_{\perp(m=n)}$ are satisfied. Hence, the probability amplitude in the 3D subspace is in the form of extended states perpendicular to the 2D effective boundary.

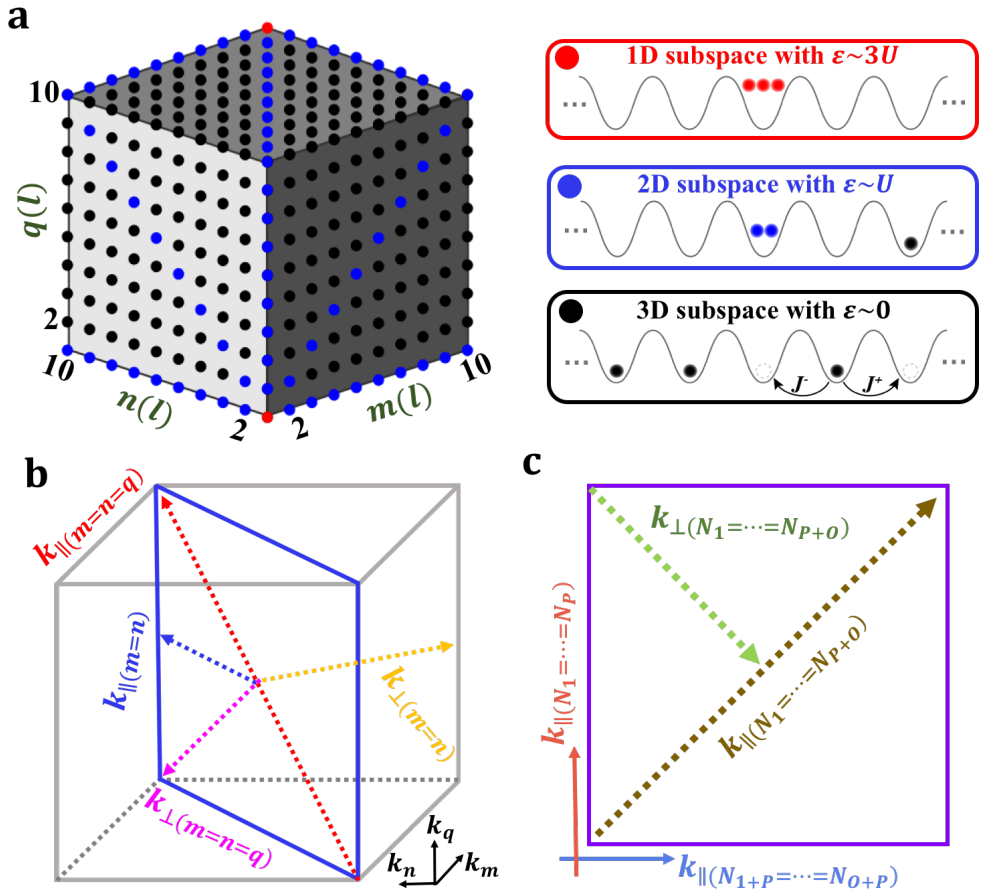

**Figure 1. Interaction-induced many-body skin effects in Hilbert space. a**, The 3D

configuration space for the probability amplitude of three bosons. Three insets enclosed by red, blue and black frames display bosonic states in three subspaces with different energies and dimensions. **b**, Illustration of wave vectors parallel/perpendicular to both 1D and 2D effective boundaries in the three-boson momentum space. **c**, The schematic diagram of wave vectors parallel/perpendicular to the effective boundary in the *N*-boson subspace possessing two bosonic clusters at different lattice sites, where the amounts for two groups of bosons are labeled by $P$ and $O$.

Then, we focus on the 2D subspace sustaining the 1D effective boundary. In this case, we derive the three-boson Bloch Hamiltonian perpendicular to the 1D effective boundary ($m = n = q$) in the 2D sub-space (with $m = n \neq q$) as:

$$H(k_{\perp(m=n=q)}) = J_{\parallel(m=n)}^{\pm} e^{\pm ik_{\parallel(m=n)}} + J_q^{\pm} e^{\pm ik_q} = J_{\perp(m=n=q)}^{\pm} e^{\pm ik_{\perp(m=n=q)}}, \quad (6)$$

where $J_{\perp(m=n=q)}^{\pm} = J_{\parallel(m=n)}^{\pm} e^{\pm ik_{\parallel(m=n=q)}} + J_q^{\mp} e^{\mp ik_{\parallel(m=n=q)}}$ are effective hopping strengths along $\pm k_{\perp(m=n=q)}$. It is noted that the symmetric hopping amplitudes along $\pm k_{\perp(m=n=q)}$ require following equations must be satisfied (see S1 in the Supplementary Materials for details):

$$\sum_{i=m,n}(J_i^+)^2 + (J_q^-)^2 = \sum_{i=m,n}(J_i^-)^2 + (J_q^+)^2, \quad J_m^+ J_n^+ = J_m^- J_n^-, \quad J_{m,n}^- J_q^+ = J_{m,n}^+ J_q^-. \quad (7)$$

With the identical principle of bosons, we note Eq. (7) is untenable, that is we always have $|J_{\perp(m=n=q)}^+| \neq |J_{\perp(m=n=q)}^-|$ in the 2D subspace. In this case, the asymmetric hopping strength perpendicular to the effective 1D boundary could make the probability amplitude of three bosons concentrate on this boundary. This phenomenon is similar to the non-Hermitian skin effect in single-particle systems with open boundaries. Hence, we call it as the non-Hermitian many-body skin effect in Hilbert space. Such a non-Hermitian induced many-body skin effect in Hilbert space represents the aggravation of two bound bosons (doublon) and an isolated boson in the lattice with periodic boundary conditions.

In the following, we generalize skin effects appeared in the three-boson Hilbert space to $N$-boson systems. The subspace possessing two bosonic clusters is considered, where $P$ bosons (labeled from $N_1$ to $N_P$) locate at one lattice site and $O$ bosons (labeled from $N_{1+P}$ to $N_{O+P}$) locate at another lattice site with $(P, O) \propto \{1, \dots, N\}$. In this case, the corresponding effective boundary could be the subspace with $(P + O)$ bosons locating at the same lattice site. Similar to the three-boson case, wave vectors parallel/perpendicular to such an effective boundary in a $N$-boson subspace are illustrated in Fig. 1c. $\boldsymbol{k}_{\|(N_1=\dots=N_P)}$ and $\boldsymbol{k}_{\|(N_{1+P}=\dots=N_{O+P})}$ represent center-of-mass wave vectors of bosonic clusters possessing $P$ and $O$ bosons, respectively. In this case, wave vectors parallel and perpendicular to the effective boundary could be expressed as $\boldsymbol{k}_{\|(N_1=\dots=N_{O+P}),\perp(N_1=\dots=N_{O+P})} = \boldsymbol{k}_{\|(N_1=\dots=N_P)} \pm \boldsymbol{k}_{\|(N_{1+P}=\dots=N_{O+P})}$.

In this $N$-boson subspace, the Bloch Hamiltonian perpendicular to the effective boundary is expressed as:

$$H\left(k_{\perp(N_1=\dots=N_{O+P})}\right) = J^{\pm}_{\perp(N_1=\dots=N_{O+P})} e^{\pm i k_{\perp(N_1=\dots=N_{O+P})}}, \tag{8}$$

where $J^{\pm}_{\perp(N_1=\dots=N_{O+P})} = J^{\pm}_{\|(N_1=\dots=N_P)} e^{\pm i k_{\|(N_1=\dots=N_{O+P})}} + J^{\mp}_{\|(N_{1+P}=\dots=N_{O+P})} e^{\mp i k_{\|(N_1=\dots=N_{O+P})}}$ are the effective hopping strengths along $\pm \boldsymbol{k}_{\perp(N_1=\dots=N_{O+P})}$. Following detailed derivations (see S2 in Supplementary Materials for details), we find that the asymmetric hopping strength $|J^{+}_{\perp(N_1=\dots=N_{O+P})}| \neq |J^{-}_{\perp(N_1=\dots=N_{O+P})}|$ along the direction of $\pm \boldsymbol{k}_{\perp(N_{1+P}=\dots=N_{O+P})}$ can only exist, when the particle numbers of two bosonic clusters are different ($P \neq O$). In this case, the interaction-induced skin effect could appear in the subspace with non-identical occupations. Physically, such a non-Hermitian induced many-body skin effect in Hilbert space represents the aggregation of bosonic clusters with different amounts of bosons. In addition, it is worthy to note that the above proposed skin states in Hilbert space only appear with periodic boundary

conditions. Under open boundary conditions, similar to the single-particle case, all correlated bosons are accumulated at physical boundaries.

To demonstrate the above predicted many-body skin effects in Hilbert space, we numerically calculate complex energy spectra and profiles of density of state for the 1D three-boson system. The parameters are set as $J^+ = -1$, $J^- = 0$, $U = 50$ and $L = 10$, respectively. Figs. 2a, 2b and 2c present the complex energy spectra for the above proposed three subspaces, respectively. To quantify the localization degree of the associated eigen-states ($\boldsymbol{\varphi}_i$), we calculate the inverse participation ratio (IPR) $IPR = \sum_{i=[1,n_d]} |\boldsymbol{\varphi}_i|^{-4}/n_d$ with $n_d$ being the number of eigen-states in the $d$-dimensional sub-space. It is noted that IPR approaches to 1 for an extended state (red), while the localized state has $IPR \sim 0$ (blue). In this case, we can see that only the eigen-states in the 2D sub-space exhibit significant localizations.

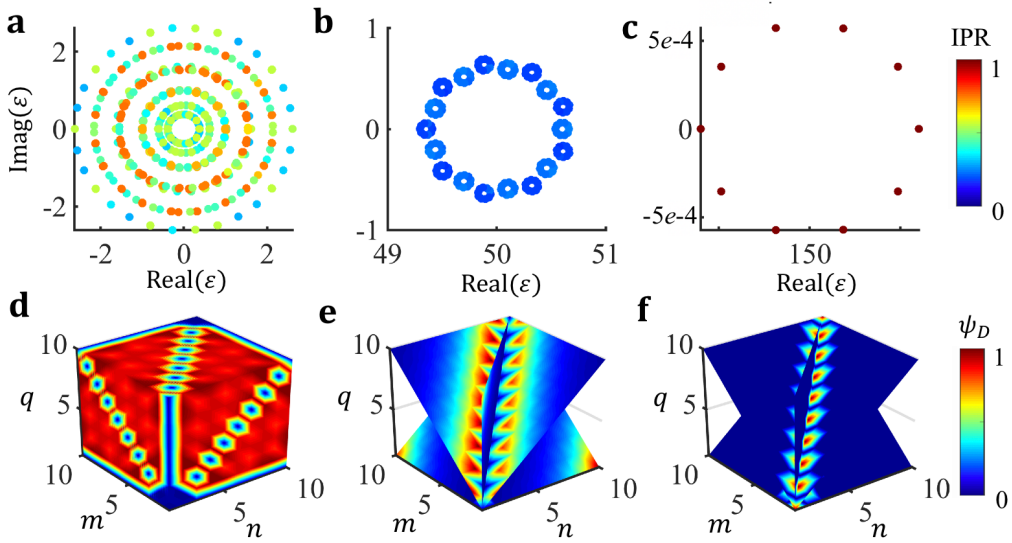

**Figure 2. Numerical results of interaction-induced many-body skin effects in Hilbert space for three bosons. a-c,** Results of the complex energy spectra of the three-boson system in three subspaces with different energies and dimensions. The color closes to blue (red) corresponds to the localized (extended) eigen-states quantified by the *IPR*. **d-f,** Illustrations of the profiles of normalized density of states in three subspaces.

Then, we calculate the normalized density of state in each sub-space $\psi_d(m,n,q) = \sum_{i=[1,n_d]} |\varphi_i(m,n,q)|^2 / \text{Max}(\sum_{i=[1,n_d]} |\varphi_i(m,n,q)|^2)$, as plotted in Figs. 2d, 2e and 2f. It is clearly shown that the density of states in 3D and 1D subspaces are both in the form of extended states. Only the density of state in the 2D sub-space is strongly localized around the 1D effective boundary. These results clearly indicate that the interaction-induced skin effect in Hilbert space of three bosons could only appear in the 2D subspace toward the 1D effective boundary. Moreover, we also calculate the complex energy spectra and corresponding density of states with different interaction strengths, and find that the localization length of many-body skin states becomes much larger when the interaction strength decreases (see S3 in Supplementary Materials for details). Except for the three-boson system, both analytical and numerical results for the four-body and five-body cases are also provided in the S4 of Supplementary Materials, where all results are consistent with our claim on the interaction-induced skin effect in Hilbert space.

In fact, the observation of theoretically predicted many-body skin states requires the control of non-reciprocal hopping and strong interactions in the quantum many-body system, which are extremely hard to be realized in experiments. In the next part, we will discuss a novel method to construct a circuit lattice to simulate the many-body skin effect in Hilbert space.

***Three-dimensional circuit lattices as a classical simulator of the interaction-induced many-body skin effect in Hilbert space.*** According to previous investigations[40-43], by mapping the many-body configuration-space to the high-dimensional lattice, the lower-dimensional Bose-

Hubbard model can be simulated by a single-particle lattice with higher dimensions. As for the above-mentioned system, the configuration-state of three bosons in Fig. 1a can be regarded as the three-dimensional lattice modes in the single-particle region. With such an analogy, the probability amplitude of the 1D three-boson model $c_{mnq}$ is directly mapped to the probability amplitude of the single particle locating at the lattice site ($m$, $n$, $q$) in the 3D space. The non-reciprocal hopping along a certain direction of the mapped 3D lattice represents the asymmetric hopping of one boson in the original 1D lattice. In addition, the on-site potential at three diagonal planes ($m=n$, $m=q$ and $n=q$) of the mapped 3D lattice can mimic the on-site interaction. In this case, the behavior of three correlated bosons in the 1D lattice can be effectively simulated by a single particle in the mapped 3D lattice.

Based on the similarity between circuit Laplacian and lattice Hamiltonian[44-50], the mapped 3D lattice can be simulated by electric circuits. As displayed in Fig. 3a, we select two representative cutting planes embedded in the 3D circuit, that are ($m$, $n$, $q$=7) with the green frame and ($m=n$, $q$) with the blue frame, to illustrate details of the circuit design. Figs. 3b and 3c present the circuit structure at these two planes. The black, blue and red circuit nodes correspond to three-boson states in 3D, 2D and 1D subspaces, respectively. To fulfill the non-reciprocal coupling of adjacent circuit nodes, the capacitor $C_1$ is parallelly connected with a one-way coupling capacitor[50], which is realized by a capacitor $C_2$ in series with a negative impedance converter with current inversion (INIC) (as shown by the top inset of Fig. 3d). The particle interaction can be simulated with the suitable grounding, as presented in Fig. 3d, where the inductor ($L_g$) is used to link each node to the ground, and the capacitor $C_U$ ($3C_U$) is selected for extra groundings on the diagonal planes (line) of blue (red) nodes. Moreover, circuit nodes at boundaries are connected end-to-end to realize the periodic boundary condition. Through the appropriate setting of grounding and connecting, we find that the circuit eigen-equation possesses the same form as the stationary Schrödinger equation of three correlated bosons as:

$$(\frac{f_0^2}{f^2}-6)V_{mnq} = -(C_1 \pm C_3)/C_1 \left(V_{m\pm1,n,q} + V_{m,n\pm1,q} + V_{m,n,q\pm1}\right) + C_U/C_1 (\delta_{mn} + \delta_{mq} + \delta_{nq})V_{mnq}$$

(8)

where $f$ is the eigen-frequency ($f_0 = 1/2\pi\sqrt{C_1 L_g}$) of the designed circuit and $V_{mnq}$ represents the voltage on the circuit node ($m, n, q$). In particular, the voltage $V_{mnq}$ corresponds to the three-boson probability amplitude $c_{mnq}$, and the eigen-energy of three bosons is directly related to the eigen-frequency of the designed circuit as $\varepsilon = f_0^2/f^2 - 6$ with other parameters being $J^{\pm} = (C_1 \pm C_2)/C_1$ and $U = C_U/C_1$. The details for the derivation of circuit eigen-equations and the correspondence to the 1D non-Hermitian Bose-Hubbard model are provided in S5 of Supplementary Materials.

Now, we perform circuit simulations using the LTSpice software to illustrate the interaction-induced skin effect in Hilbert space. Here, the value of $C_1, C_2, C_U$ and $L_g$ are taken as 1$nF$, 1$nF$, 10$nF$ and 3.3$uH$ (the same parameters of these elements are used below). To test whether skin effects exist in the 3D sub-space toward the 2D effective boundary, impedance responses of different circuit nodes along the direction of $\boldsymbol{k}_{\perp(m=n,q)}$ should be analyzed. Here, we select three circuit nodes in the 3D sub-space along $\boldsymbol{k}_{\perp(m=n,q=1)}$, where positions of these circuit nodes are labeled by (8,3,1), (7,4,1) and (6,5,1), as shown by black dots in Fig. 3a. Figs. 3e, 3f and 3g display numerical results of the calculated impedance (with respect to the ground) of these circuits nodes in three frequency ranges, which correspond to eigen-energies of three bosons in different subspaces ($\varepsilon \sim 3U, U$ and $0$), respectively. It is clearly shown that impedance peaks of three circuit nodes all appear in the frequency range from 0.8MHz to 1.5MHz, being matched to the corresponding eigen-energy of three bosons in the 3D sub-space with $\varepsilon\sim 0$. Moreover, we note that peak values of these circuit nodes are nearly identical. This indicates that no skin effect appears in the 3D sub-space toward the 2D effective boundary, which is accord with the theoretical prediction.

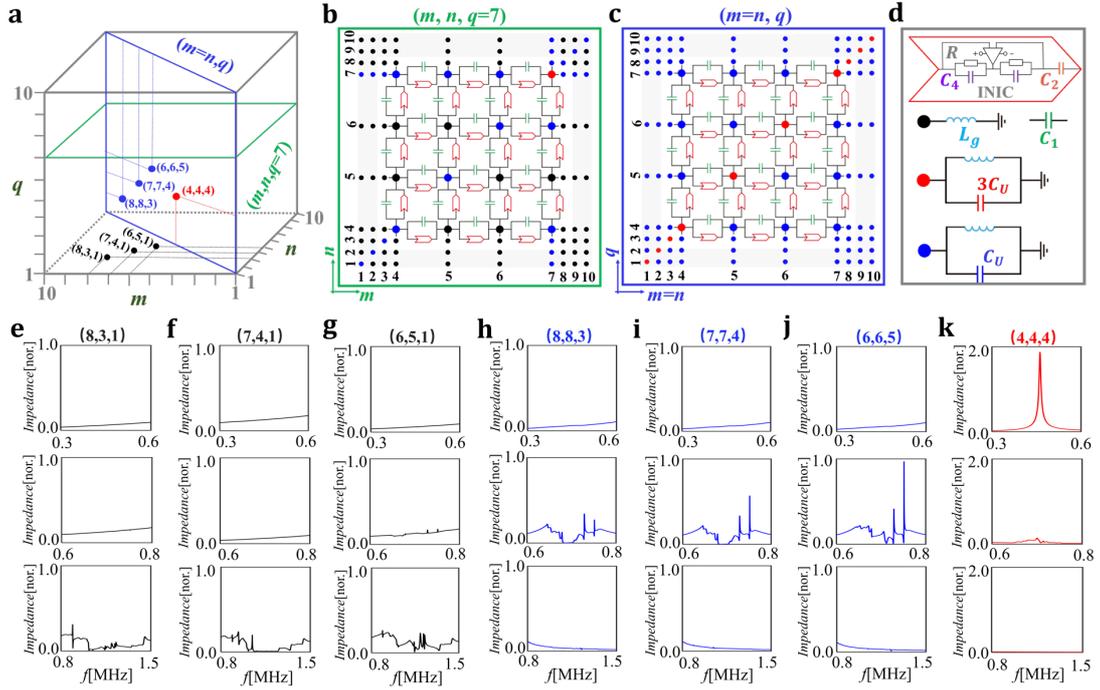

**Figure 3. The schematic diagram and simulation results of designed 3D electric circuits.** **a**, The diagram of two representative cutting planes embedded in the 3D circuit, that are ($m$, $n$, $q$=7) with the green frame and ($m$=$n$, $q$) with the blue frame. Black and blue dots mark circuit nodes along the direction of $k_{\perp(m,n,q=1)}$ and $k_{\perp(m=n,q)}$. The red dot marks the circuit node on the diagonal line at (4,4,4). **b** and **c** present the designed electric circuit at two cutting planes ($m$, $n$, $q$=7) and ($m$=$n$, $q$), respectively. **d**, The ground setting of different circuit nodes and the structure of the one-way coupling capacitor with $C_4$=10$nF$ and $R$=300Ω. **e-g**, The impedance responses of three circuit nodes locating in the 3D sub-space along $k_{\perp(m,n,q=1)}$. **h-j,** The impedance responses of three circuit nodes in the 2D sub-space along $k_{\perp(m=n=q)}$. **k**, The simulated impedance responses of a circuit node in the 1D sub-space at (4,4,4).

Next, the impedance responses of different circuit nodes along the direction of $k_{\perp(m=n=q)}$ are calculated to prove the existence of skin effects in the 2D subspace toward the 1D effective boundary. Figs. 3h, 3i and 3j present simulated impedance responses of three selected circuit nodes, which locate at (8,8,3), (7,7,4) and (6,6,5) as marked by blue dots in Fig. 3a. It is clear shown that impedance peaks appear at the frequency around 0.69MHz, which is matched to the eigen-energy of three bosons in the 2D sub-space ($\varepsilon \sim U$). Importantly, it is clearly shown that peak values get increased with circuit nodes approaching to the effective 1D boundary. This phenomenon demonstrates the appearance of skin states in the 2D sub-space toward the 1D effective boundary. Finally, the impedance response of a circuit node in the 1D subspace at (4,4,4) is calculated, as shown in Fig. 3k. We find that a good consistence is obtained between the frequency of the impedance peak (around 0.46MHz) and the three-boson eigen-energy in the 1D sub-space ($\varepsilon \sim 3U$).

Based on the above impedance simulations, the interaction-induce many-body skin states are clearly verified. In fact, there is another way to prove the existence of interaction-induced many-body skin effects by recovering the circuit admittance spectrum[50]. According to such a method, the recovered admittance eigen-values and eigen-states of our designed 3D circuit clearly support the existence of skin effects in Hilbert space. See detailed results in S6 of Supplementary Materials.

***Experimental observation of interaction-induced many-body skin effects in Hilbert space by electric circuits.*** To experimentally observe the interaction-induced many-body skin effects in Hilbert space, the 3D electric circuit possessing same parameters with that considered in simulations is fabricated. The photograph image of the circuit sample is presented in Fig. 4a. Here, a totally ten printed circuit boards (PCBs) are applied with a single PCB containing 10×10 nodes in the *mn*-plane. The enlarged views for the front and back sides of a single PCB are shown in Figs. 4b and 4c. The adjacent circuit nodes along each axis are connected through the capacitor $C_1$ (framed by the black circle) and a parallelly connected one-way capacitor, which is realized by connecting a capacitor $C_2$ (enclosed by the red circle) in series with a negative INIC (enclosed by the blue block). The grounding capacitor $C_U$ on the diagonal plane is marked by the green block in the front side of the sample, and the grounding inductor $L_g$ is marked by the yellow block in the back side. It is worthy to note that the tolerance of circuit elements is only 1% to avoid the detuning of circuit responses. Details of the sample fabrication is provided in Methods.

We use Wayne kerr precision impedance analyzer to measure the impedance response of circuit nodes as a function of the driving frequency. Figs. 4d, 4e and 4f present the measure impedances of three circuit nodes along the direction of $\boldsymbol{k}_{\perp(m=n,q=1)}$ in the 3D sub-space, where the selected circuit nodes are the same to those used in Figs. 3e, 3f and 3g. We note that the frequency range of measured impedance peaks is accord with simulation results, and peak values of these circuit nodes are nearly identical. These measured results clearly demonstrate that there is no skin effect in the 3D sub-space toward the 2D effective boundary.

In Figs. 4g, 4h and 4i, we plot the measured results of impedance responses for three circuit nodes along the direction of $\boldsymbol{k}_{\perp(m=n=q)}$ in the 2D-subspace (the same to circuit nodes considered in Figs. 3h, 3i and 3j). It is shown that significant impedance peaks only exist in the frequency range around 0.69MHz, and the peak value gets increased with circuit nodes approaching to the 1D effective boundary. These experimental results are also in a good accordance with simulation results, proving the appearance of interaction-induced skin effect in the 2D sub-space toward the 1D effective boundary. The wider peaks compared to that of numerical results are mainly due to the larger lossy effect, resulting from the resistive loss of linking wires and the low Q-factor of the applied inductor. Lastly, Fig. 4j displays the measured impedance of a circuit node at (4,4,4) in the 1D subspace. It is shown that the peak value appears at the frequency around 0.46MHz, which is also consistent with the simulation result.

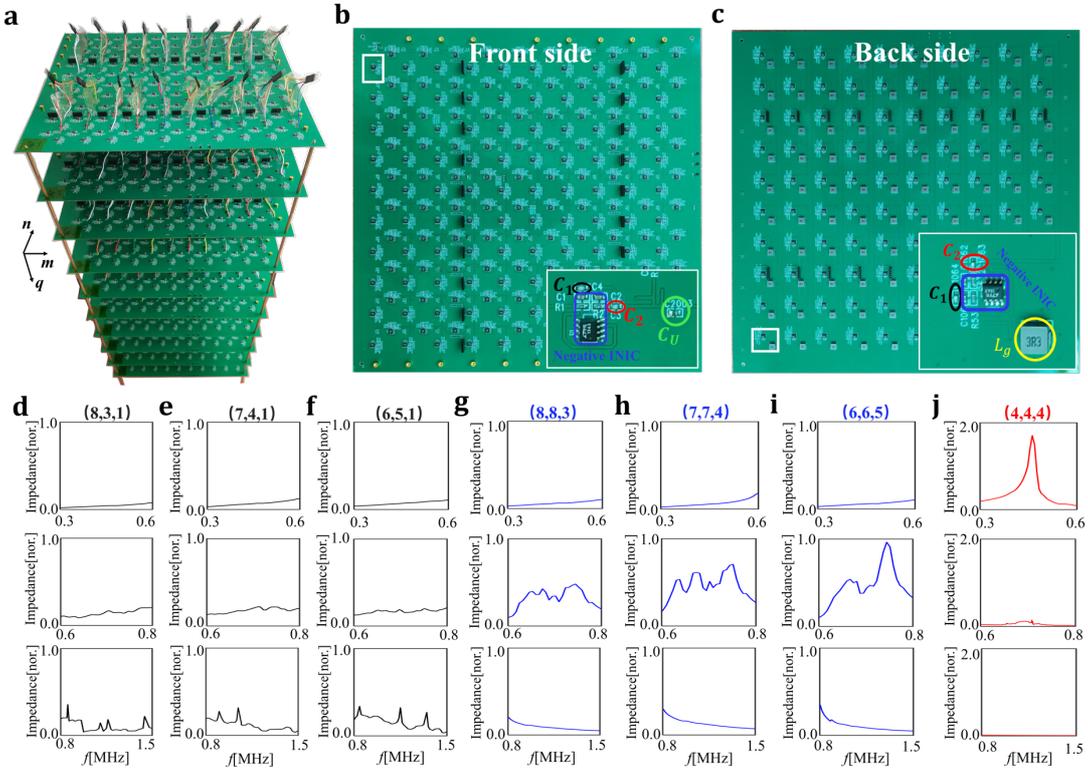

**Figure 4. Experimental results for observing interaction-induced many-Body skin effects in Hilbert Space. a**, The photograph image of the fabricated 3D circuit. **b** and **c**, The enlarged views for the front and back sides of a single PCB. **d-f**, The measured impedance responses of three circuit nodes located in the 3D sub-space along $k_{\perp(m=n,q=1)}$. **g-i**, The measured impedances of three circuit nodes located in the 2D sub-space along $k_{\perp(m=n=q)}$. **j**, The measured impedances of a circuit node in the 1D sub-space at (4,4,4).

**Discussion and conclusion.**

In conclusion, we have theoretically proposed and experimentally demonstrated that the strong interaction can induce novel non-Hermitian skin effects in the many-body Hilbert space. The direct analytical derivations indicate that the skin boundary should locate in the subspace sustaining clusters with different amounts of correlated bosons. Using three-boson system as the illustration, we map the corresponding eigenstates to modes of the designed circuit lattice to simulate the strong interaction-induced many-body skin effects. Both circuit simulations and measurements give a clear evidence for the appearance of non-Hermitian many-body skin states in Hilbert Space. It is also interesting to investigate other properties of many-body skin effects in Hilbert space based on the designed circuit simulator, such as critical phenomena and the competition with disorder-induced localizations. Our proposal provides a flexible platform to investigate and visualize many interesting phenomena related to the particle interaction and non-Hermitian physics, and could have potential applications in the field of the novelty electronic signal control.

**Methods.**

**Sample fabrications.** We exploit the electric circuits by using PADs program software, where the PCB composition, stack up layout, internal layer and grounding design are suitably engineered. Here, the well-designed PCB possesses totally eight layers to arrange the intralayer and interlayer site-couplings. It is worthy to note that the ground layer should be placed in the gap between any two layers to avoid their coupling. Moreover, all PCB traces have a relatively large width (0.5mm) to reduce the parasitic inductance and the spacing between electronic devices is also large enough (1.0mm) to avert spurious inductive coupling. The SMP connectors are welded on the PCB nodes for the signal input and measurement. To ensure the performance of designed electric circuits, the tolerance of circuit elements should be as low as possible. For this purpose, we use WK6500B impedance analyzer to select circuit elements with high accuracy (the disorder strength is only 1%).

**Acknowledgements**

This work was supported by the National key R & D Program of China under Grant No. 2017YFA0303800 and the National Natural Science Foundation of China (No.91850205 and No.61421001).